\documentclass{moriond}

\usepackage{amsthm,amsmath,amssymb,mathrsfs}
\usepackage[super]{cite}

\def\be{\begin{equation}}
\def\ee{\end{equation}}
\def\bea{\begin{eqnarray}}
\def\eea{\end{eqnarray}}

\newcommand{\Photo}{\includegraphics[height=35mm]{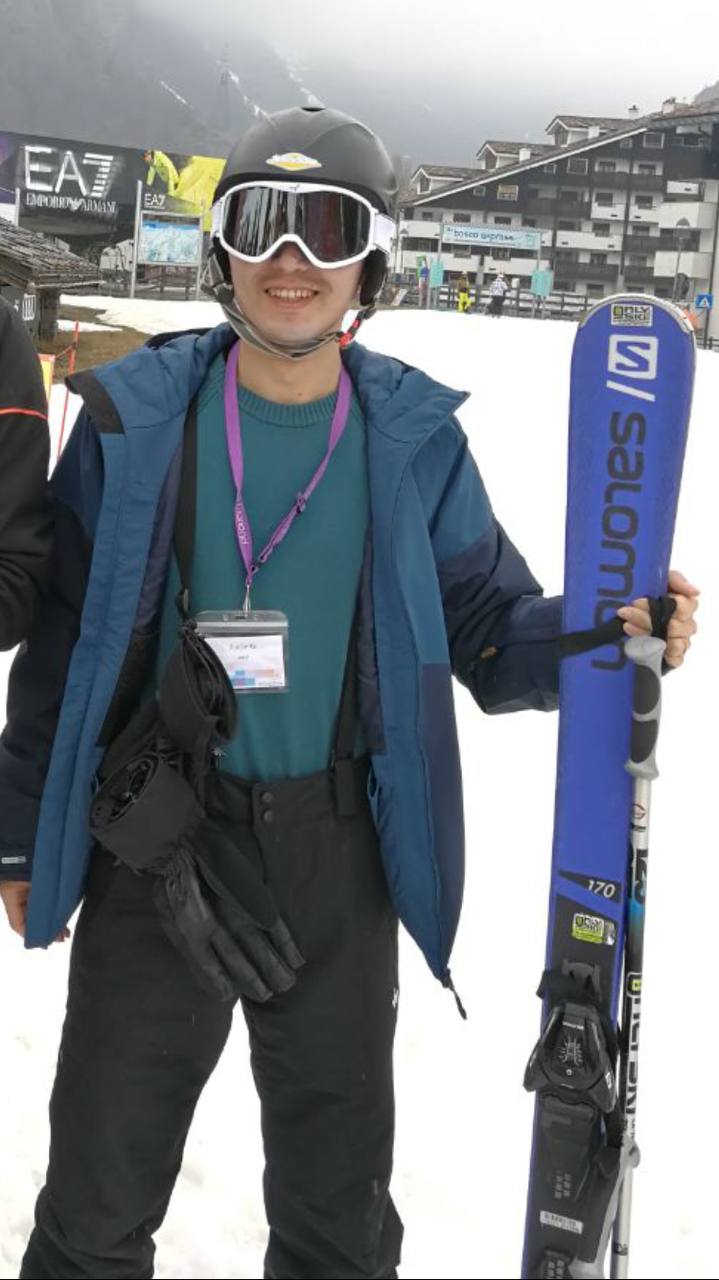}}

\begin{document}
\vspace*{4cm}

\title{Neutrino forces and experimental probes} 

\author{Xun-Jie Xu}

\address{Institute of High Energy Physics, Chinese Academy of Sciences, Beijing 100049, China}

\author{in collaboration with \\ Rupert Coy, Mijo Ghosh, Yuval Grossman,
	Walter Tangarife, and Bingrong Yu}

\maketitle

\abstracts{ \vspace{3em} Neutrinos as almost massless particles
	could mediate long-range forces, known as neutrino forces. In this
	talk, I will introduce some theoretical aspects of neutrino forces,
	including why the potential of a neutrino force has the $1/r^{5}$
	form and how it may vary under different circumstances. Experimental
	probes and possible implications for cosmology are also briefly discussed.
}

\section{Introduction}

Neutrino forces are an interesting SM prediction which has not yet
been observed successfully. In this talk, I am going to introduce
some fun theoretical aspects and a few interesting ideas about how
neutrino forces might be probed in experiments.

As is well known, whenever there is a massless particle propagating
freely, it always mediates a long-range force. This is true for the
graviton and the photon. What about neutrinos? Neutrinos can be massless.
Can they mediate long-range forces? The answer is yes. But since neutrinos
are fermions, you have to put a pair of neutrinos between two test
particles so that the spin of each test particle is unchanged. The
force generated in this way is called a neutrino force.

\begin{figure}[h]
	\centering
	
	\includegraphics[width=0.3\textwidth]{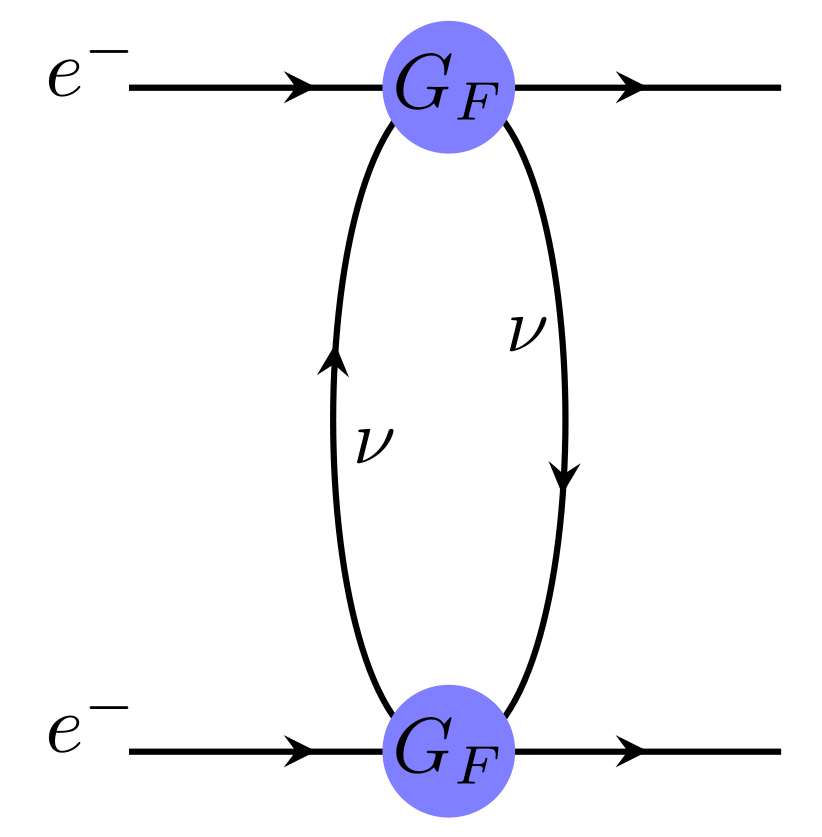}\caption{Feynman diagram for the SM neutrino force.\label{fig:feyn}}
\end{figure}

The idea of neutrino forces has been there for quite a long time.
In the 1930s, it was proposed as a possible nuclear force~\cite{manypapers}.
And in the 1960s, it was quantitative computed for the first time
by Feinberg and Sucher~\cite{Feinberg:1968zz}. At almost the same
time, Feynman discussed neutrino forces in his famous lectures~\cite{Feynmangravitation}.
The interesting point is that Feynman demonstrated how one could make
neutrino forces behave like gravity by adding a third body that exchanges
neutrinos with the other two test bodies. In the 1990s, there was
a debate stemming from an argument that neutrino forces could lead
to instability of neutron stars~\cite{Fischbach:1996qf}, which was
later criticized by many authors~\cite{Smirnov:1996vj,Abada:1996nx,Kachelriess:1997cr,Kiers:1997ty,Abada:1998ti,Arafune:1998ft}. 

\section{Neutrino forces in the vacuum}

The effective potential of a neutrino force, e.g.~between two electrons,
has the following form~\cite{Feinberg:1968zz,Feinberg:1989ps,Hsu:1992tg}:
\begin{equation}
	V_{ee}(r)=\left(2\sin^{2}\theta_{W}+\frac{1}{2}\right)^{2}\frac{G_{F}^{2}}{4\pi^{3}}\frac{1}{r^{5}}\thinspace.\label{eq:}
\end{equation}
It is proportional to $G_{F}^{2}/r^{5}$, which can be understood
from dimensional analysis. The diagram in Fig.~\ref{fig:feyn} involves
two $G_{F}$'s so it has to be proportional to $G_{F}^{2}$, which
is of dimension $[E]^{-4}$. The potential is a dimension-one quantity.
To balance the dimension, $1/r^{5}$ has to be added. If we replace
the neutrino by a massless scalar, we would get $1/r^{3}.$ This can
also be easily obtained from dimensional analysis.\footnote{In fact, in {\it Feynman Lectures on Gravitation}~\cite{Feynmangravitation},
	the two-body neutrino force derived by Feynman has exactly the $1/r^{3}$
	form because he used a scalar-like propagator.}\\

Let us get a sense of how weak the neutrino force is by comparing
it to gravity. The former is proportional to $1/r^{5}$ while the
latter is proportional to $1/r$. So at a sufficiently short distance,
we expect that the neutrino force should be stronger than the gravity.
At long distances with $r>{\cal O}(10^{-8})$ cm \footnote{$10^{-8}$ cm  is only for forces between two atoms. Between two electrons, the neutrino force is of the same order of magnitude but gravity is much weaker.}, gravity starts to
dominate. Depending on whether the distance scale is above or below
$10^{-8}$ cm (corresponding to the energy scale of keV), I categorize
possible experimental probes into short- and long-range ones. 

For short-range probes, there have been proposals using atomic and
nuclear spectroscopy and atomic parity violation~\cite{Stadnik:2017yge,Ghosh:2019dmi}.
This typically requires solving the Schr\"odinger equation. However,
if the $1/r^{5}$ potential is plugged into the Schr\"odinger equation,
one immediately encounters a problem in solving the Schr\"odinger
equation. The singularity at $r=0$ makes the calculation badly divergent.
It has been well known (see Ref.~[\citen{Frank:1971xx}] for
a review) that for potentials more divergent than $1/r^{2}$ at $r=0$,
one cannot get a consistent treatment in quantum mechanics. 

\begin{figure}[h]
	\centering
	
	\includegraphics[width=0.8\textwidth]{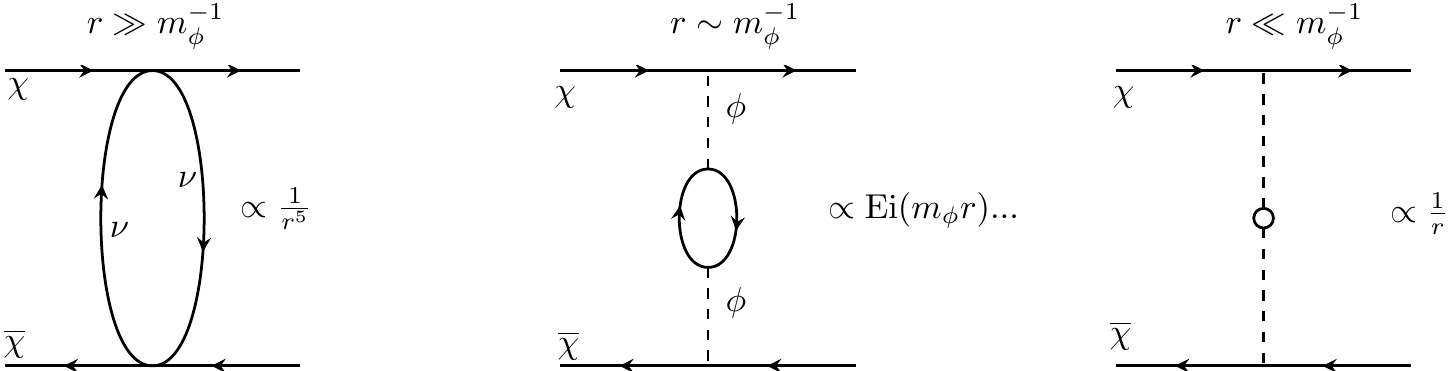}
	\caption[]{A diagrammatic explanation of how the potential of a neutrino force
		vary from $1/r^{5}$ to $1/r$. Figure taken from Ref.~[\citen{Xu:2021daf}].		}
	\label{fig:feyn-2}
\end{figure}

So we may ask whether the $1/r^{5}$ potential is always valid down
to arbitrarily small $r$. The answer is no. In our first work on
neutrino forces~\cite{Xu:2021daf}, we address exactly this issue.
Our conclusion is that, as $r$ approaches zero, the potential changes
from $1/r^{5}$ to $1/r^{4}$, $1/r^{2}$ and eventually to $1/r$.
To demonstrate that, one can insert a mediator with mass $m_{\phi}$
in this effective vertex---see Fig.~\ref{fig:feyn-2}. At a short
distance comparable to $m_{\phi}^{-1}$, the effective vertex opens
up, leading to a relatively complicated result involving the exponential
integral function, $\text{Ei}(m_{\phi}r)$. At very short distances,
which correspond to very high energy scales, the Ei function reduces
to $1/r$.

For long-range probes, they necessarily require precision test of
gravity, which is stronger than neutrino forces at $r>{\cal O}(10^{-8})$
cm. In the past two decades, the experimental progress in testing
the properties of gravity such as the inverse square law and the weak/strong
equivalent principles is remarkable. For example, assuming the presence
of a new long-range force that is not proportional to the inertial
mass, even a very small contribution of $\delta V/V_{{\rm gravity}}\sim10^{-16}$
at $\langle r\rangle\sim6400\ \text{km}$ can be excluded---see Tab.~\ref{tab:exp_list}.
However, this is still not enough to compensate the difference between
$1/r$ and $1/r^{5}$. 

\begin{table*}
	\centering
	
	\begin{tabular}{ccccc}
		\hline 
		exp & $\delta V/V_{{\rm gravity}}$ & $\langle r\rangle$ &   \tabularnewline
		\hline 
		Washington2007 & $3.2\times10^{-16}$ & $\sim6400$ km &   \tabularnewline
		Washington1999 & $3.0\times10^{-9}$ & $\sim0.3$ m &   \tabularnewline
		Irvine1985 & $0.7\times10^{-4}$ & $2-5$ cm &   \tabularnewline
		Irvine1985 & $2.7\times10^{-4}$ & $5-105$ cm &   \tabularnewline
		Wuhan2012 & $10^{-3}$ & $\sim2$ mm &  \tabularnewline
		Wuhan2020 & $3\times10^{-2}$ & $\sim0.1$ mm &  \tabularnewline
		Washington2020 & $\sim1$ & $52$ $\mu$m &   \tabularnewline
		Future levitated optomechanics & $\sim10^{4}$ & 1 $\mu$m &  \tabularnewline
		\hline
	\end{tabular}
	
	\caption[]{\label{tab:exp_list} Sensitivities of long-range force search experiments, taken from Ref.~[\citen{Ghosh:2022nzo}]. }
\end{table*}

\section{Neutrino forces in strong neutrino backgrounds}

Is it possible to enhance neutrino forces at long distances so that
they do not decrease as $1/r^{5}$? Yes, only if there is a strong
neutrino background. 

\begin{figure}[h]
	\centering
	
	\includegraphics[width=0.6\textwidth]{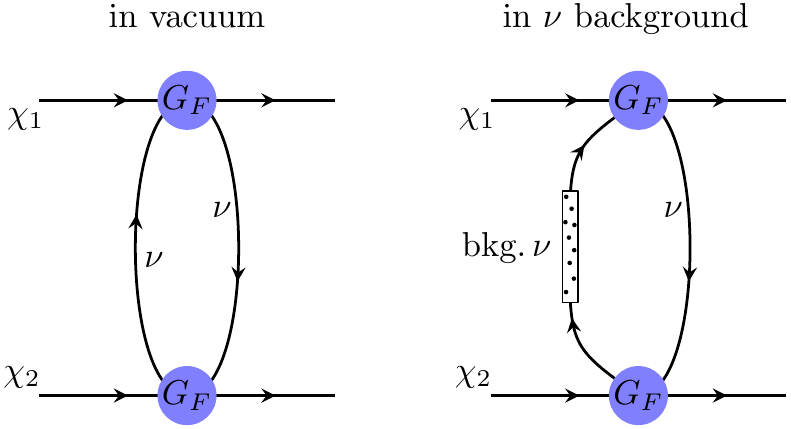}\caption{Feynman diagrams for neutrino forces in the vacuum (left) and in a
		neutrino background (right). Figure taken from Ref.~[\citen{Ghosh:2022nzo}]. \label{fig:feyn-3}}
\end{figure}

With a strong neutrino background, it is possible to have a neutrino
from the background absorbed by one test body, which then passes a
virtual neutrino to  the other test body. Then the latter returns
a neutrino to the background. This is explained by the right Feynman
diagram in Fig.~\ref{fig:feyn-3}.

This explanation might seem a bit counter-intuitive from the perspective
of point-like particles with fixed spacetime coordinates and momenta.
But it is based on the well-established theory of finite-density QFT,
which suggests that the propagator of a fermion $f$ in a finite-density
background should be modified as follows:
\begin{equation}
	S_{f}(k)=\left(k\cdot\gamma+m_{f}\right)\cdot\left[\frac{i}{k^{2}-m_{f}^{2}}-2\pi\delta\left(k^{2}-m_{f}^{2}\right)\Theta\left(\pm k^{0}\right)n_{f\pm}\left(\mathbf{k}\right)\right],\label{eq:S}
\end{equation}
where $\Theta$ is the Heaviside theta function, and $n_{f\pm}\left(\mathbf{k}\right)$
denote the number densities of background $f$ and $\overline{f}$
(anti-fermion) in the phase space. Without the background (i.e.,
$n_{f\pm}=0$), we recover the familiar propagator $S_{f}(k)=i/(k\cdot\gamma-m_{f})$
in the vacuum. 

Using the modified propagator to compute neutrino forces in the background
of a monochromatic neutrino flux, we obtain the following potential~\cite{Ghosh:2022nzo}:
\begin{equation}
	V\approx-\frac{1}{\pi}G_{F}^{2}\Phi_{0}E_{\nu}\frac{1}{r}\left\{ \cos^{2}\left(\frac{\alpha}{2}\right)\cos\left[\left(1-\cos\alpha\right)E_{\nu}r\right]+\sin^{2}\left(\frac{\alpha}{2}\right)\cos\left[\left(1+\cos\alpha\right)E_{\nu}r\right]\right\} ,\label{eq:V-alpha}
\end{equation}
where $\Phi_{0}$ is the neutrino flux, $E_{\nu}$ is the neutrino energy, and $\alpha$ denotes the angle
between the flux and the vector $\mathbf{r}$. In the $\alpha=0$
limit, the potential is simply 
\begin{equation}
	V\approx-\frac{1}{\pi}G_{F}^{2}\Phi_{0}E_{\nu}\frac{1}{r}\thinspace,\label{eq:V-1}
\end{equation}
which implies a huge enhancement---the potential is now proportional
to $1/r$ instead of $1/r^{5}$. 

However, one should note that Eq.~\eqref{eq:V-alpha} is highly oscillatory
if $\alpha\neq0$ and $E_{\nu}r\gg1$. So in realistic scenarios where
$\alpha$ cannot be perfectly zero and $E_{\nu}$ is typically much
larger than $r^{-1}$, the $1/r$ behavior can be easily smeared out.
This was also pointed out in Ref.~[\citen{Blas:2022ovz}]. Nevertheless,
if future experiments could achieve measurements with very small $\alpha$
satisfying~\cite{Ghosh:2022nzo}
\begin{equation}
	\alpha^{2}\lesssim\frac{\pi}{\Delta(E_{\nu}r)}\thinspace,\label{eq:alpha}
\end{equation}
then the strong enhancement and the $1/r$ form could still be probed.

\section{Neutrino forces and dark matter}

\begin{figure}[h]
	\centering
	
	\includegraphics[width=0.465\textwidth]{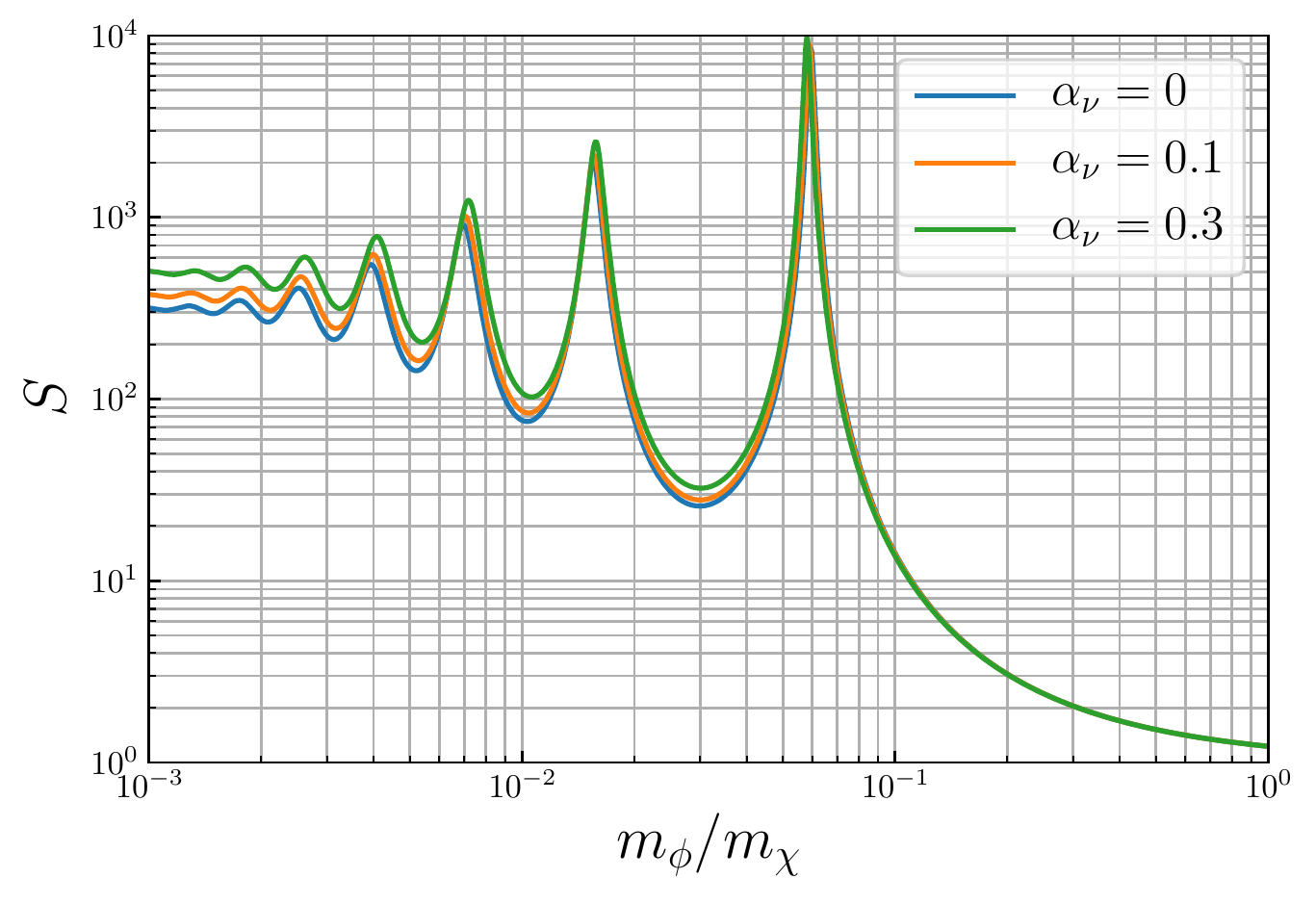}\includegraphics[width=0.48\textwidth]{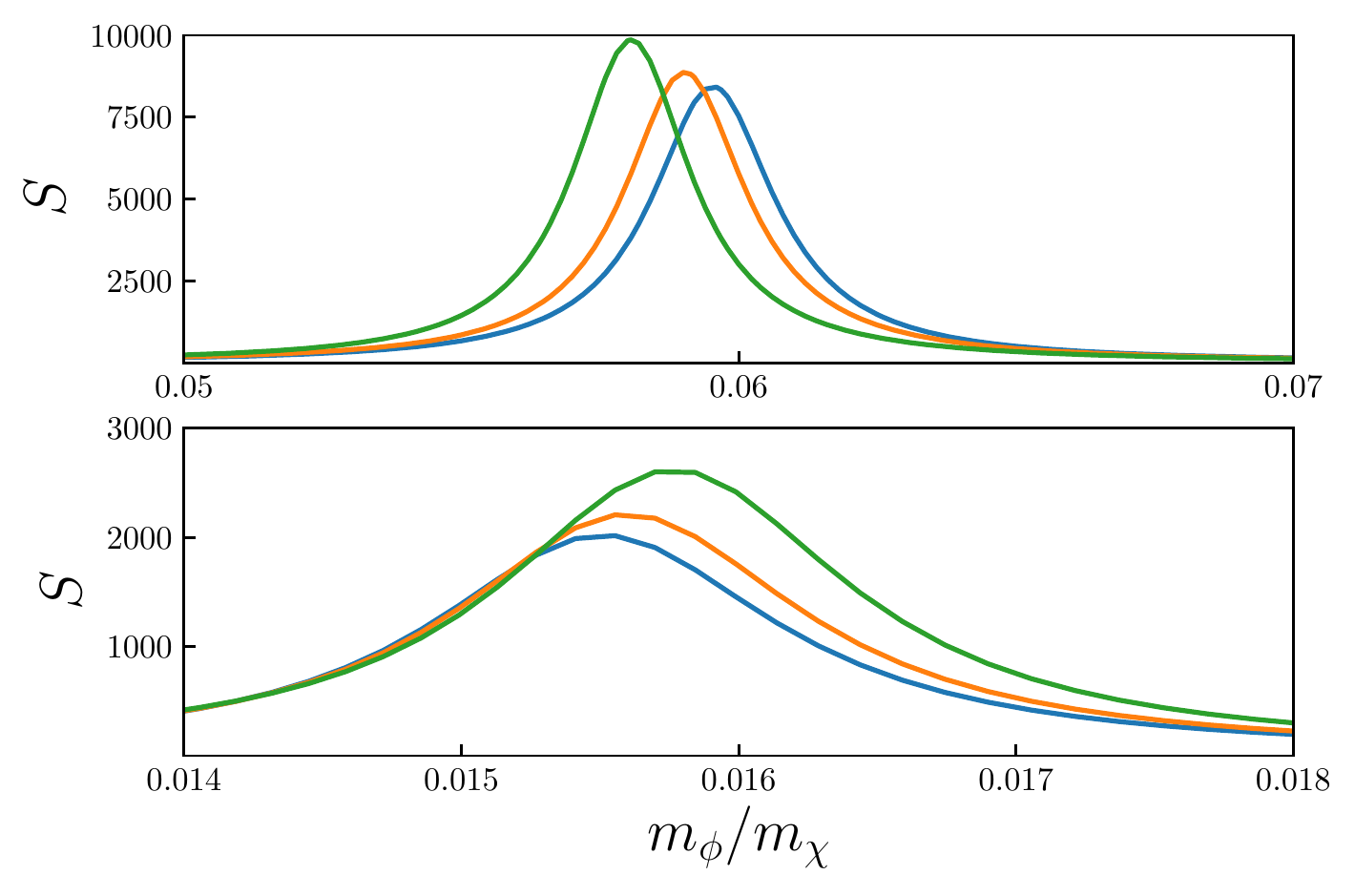}\caption{The Sommerfeld enhancement in DM annihilation modified by neutrino
		forces. The left panel shows the entire mass range whereas the right
		panels show zoom-in views of the two highest peaks. For further details,
		see Ref.~[\citen{Coy:2022cpt}]. \label{fig:DM}}
\end{figure}

Neutrino forces might also have important implications for cosmology
and dark matter (DM). Neutrinos, if interacting with DM, can mediate
a long-range force between DM particles. It has been shown that~\cite{Orlofsky:2021mmy},
due to the long-range feature of neutrino forces, the cross section
of DM self-scattering could be enhanced when the scattering is soft.
Therefore, neutrino forces could be responsible for DM self-interactions
used to resolve problems in the small-scale structure formation.

Another example is that neutrino forces could modify the so-called
Sommerfeld enhancement in DM annihilation~\cite{Coy:2022cpt}. In
general, long-range attractive forces could substantially enhance
the annihilation rate of slow-moving DM particles. This is  known
as the Sommerfeld enhancement. As is shown in Fig.~\ref{fig:DM},
the Sommerfeld enhancement can be significantly modified in the presence
of DM-neutrino interactions. Focusing on e.g.~the highest peaks of
the Sommerfeld enhancement, the heights can be changed up to $\sim60\%$
with a reasonably strong DM-neutrino coupling, and the positions of
these peaks can be shifted as well. 

\section{Summary}

In this talk, I have introduced some interesting features of neutrino
forces. In the vacuum, the potential of a neutrino force  decreases as
$1/r^{5}$, which can be easily obtained from dimensional analysis.
If $r$ is very small, then the $1/r^{5}$ form would be invalid and
we have discussed how the potential would vary in the short-range
limit ($1/r^{4}$, $1/r^{2}$, $1/r$). On the other hand, if there
is a strong neutrino background, then neutrino forces can  behave
very differently. In particular, in the small-$\alpha$ limit, the
force could be behave as  $1/r$ even in the long-range regime. So
theoretically, a strong neutrino background could significantly enhance
neutrino forces. In cosmology, neutrino forces could affect DM annihilation
rates via the Sommerfeld enhancement. 

So far neutrino forces are still yet to be experimentally probed,
despite some ideas or proposals involving parity violation effects, atomic spectroscopy, torsion balance devices, etc. Therefore,
we are looking forward to the possibility of probing neutrino forces
in the future with more advanced technologies and novel ideas.

\section*{Acknowledgments}

I would like to the organizers of the Rencontres de Moriond for the great opportunity to learn skiing (and, of course, to enjoy physics as well).
This work is supported in part by the National Natural Science Foundation
of China under grant No. 12141501.

\section*{References}
\bibliography{ref}

\end{document}